\documentclass[a4paper,11pt]{article}  
\usepackage{graphicx,amsmath,amssymb,dsfont,bm}
\usepackage{color}
\usepackage{enumitem}
\usepackage{boldline}
\usepackage{geometry}
\usepackage{authblk}
\usepackage[noadjust]{cite}

\title{\bf Trading strategies for stock pairs regarding to the cross--impact cost}

\author{Shanshan Wang\thanks{shanshan.wang@uni-due.de} }
\affil{\textit{Fakult\"at f\"ur Physik, Universit\"at Duisburg--Essen, Lotharstra\ss e 1, 47048 Duisburg, Germany}}

\date{\today}
\providecommand{\keywords}[1]{\textbf{Keywords:} #1}
 
\begin{document}
\maketitle

\begin{abstract}
We extend the framework of trading strategies of Gatheral [2010] from single stocks to a pair of stocks. Our trading strategy with the executions of two round--trip trades can be described by the trading rates of the paired stocks and the ratio of their trading periods. By minimizing the potential cost arising from cross--impacts, \textit{i.e.} the price change of one stock due to the trades of another stock, we can find out an optimal strategy for executing a sequence of trades from different stocks. We further apply the model of the strategy to a specific case, where we quantify the cross--impacts of traded volumes and of time lag with empirical data for the computation of costs. We thus picture the influence of cross--impacts on the trading strategy. 
\end{abstract}


\section{Introduction}
\label{sec1}

In financial markets, a buy or sell market order may lead to the subsequent price change. This change on average is termed price impact, which has been extensively studied over the past decades~\cite{Kyle1985,Torre1997,Lillo2003,Bouchaud2004,Bouchaud2010}. Due to the obscure origin in market microstructure~\cite{Bouchaud2004, Farmer2004, Bouchaud2010b,Farmer2007,Gomes2015} and the important role in the cost estimation for trading~\cite{Almgren2005,Freyre2004}, the price impact attracts much attention from academic and industry researchers. Most previous studies focus on single stocks, where the price impact is termed price self--impact. For fixed order size, the self--impact for single transaction is transient~\cite{Bouchaud2004} due to the market resilience, \textit{i.e.}, the new coming limit orders as the counterparty of the market orders provide the market liquidity and reverse the price gradually. Without a time lag, the self--impact behaves as a power--law function of order size~\cite{Lillo2003,Lillo2004,Gabaix2003,Plerou2004}. Therefore, both the order size and the time lag work on the price self--impact, leading to a temporary and a permanent component of impacts. Recently, the studies~\cite{Wang2016a,Wang2016b,Benzaquen2016} manifest the presence of the price impact across different stocks, which is termed price cross--impact. They provide the empirical evidence for the early discussions on the cross--impact in model--based studies~\cite{Hasbrouck2001,Almgren2001,Pasquariello2013,Cartea2015}. The cross--impact is very likely to be found in portfolios, as many stocks in a portfolio are traded in a very similar way, prompting the correlation of order flow across different stocks. 

The price change due to a market order will lead to an extra cost for trading. To reduce the transaction cost, the traders are unlikely to submit extremely large market orders directly, as these orders will consume the volumes at each price level in the order book and shift the price to a higher or a lower level immediately. Instead, they prefer to execute a sequence of small orders, which are split from a large order. By this way, it largely lowers the cost from the price impact. To reduce the transaction cost as much as possible, therefore, a lot of optimal trading strategies are proposed ~\cite{Almgren2001,Obizhaeva2013,Gatheral2010,Gatheral2013,Alfonsi2014,Alfonsi2016,Curato2017}. Most of these strategies, however, focus on the self--impact cost for single stocks, ignoring the cross--impact cost between stocks. Only very few of execution strategies take the cross--impact into account~\cite{Almgren2001,Cartea2015,Schneider2016}. In particular, Almgren and Chriss [2001] consider the optimal execution for portfolio transactions by minimizing a combination of volatility risk and transaction costs, where the permanent and temporary impacts are set to be linear in the rate of trading. Cartea \textit{et al.} [2015] construct an optimal execution strategy for liquidating a basket of assets whose price are co--integrated. In their model, they assume linear temporary and permanent price impacts in the speed of trading as well. Recently, Schneider and Lillo [2016] extend the framework of trading strategies from Gatheral [2010], \textit{i.e.}, Ref.~\cite{Gatheral2010}, to multiple assets, and discuss the possible constraints on the shape and size of the cross--impact. On condition of absence of dynamical arbitrage, they found the cross--impact must be odd and linear function of trading intensity and symmetric for a pair of stocks. However, empirical studies show a nonlinear impact function of order size either in single stocks or across multiple stocks~\cite{Lillo2003,Potters2003,Wang2016c}. 

Our study aims to construct a trading strategy regarding to the cross--impact cost, and reveal the influence of cross--impacts on the trading strategy. We thus extend the framework of trading strategies for single stocks from Gatheral [2010] to a pair of stocks, where the trading price is generated by the price self-- and cross--impact model from Ref.~\cite{Wang2016c}. Our trading strategy for a pair of stocks results from three parameters, \textit{i.e.}, the rate of trading for each stock, and the ratio of trading periods of two stocks. By minimizing the cross--impact cost, we thus can obtain an optimal trading strategy. In view of the previous empirical results~\cite{Lillo2003,Bouchaud2004,Wang2016c} and our empirical analysis,  when applying this strategy to a specific case, we employ the power--law impact functions of time lag and of order size to compute the costs.  

This paper is organized as follows. In Sec.~\ref{sec2}, making use of the price impact model with both self-- and cross--impacts, we construct a trading strategy in terms of the cross--impact cost. In Sec.~\ref{sec3}, we apply the trading strategy to a pair of stocks, where we quantify the cross--impacts of traded volumes and of time lag for the need of computation, resulting in the visible strategy depending on three parameters. We conclude our results in Sec.~\ref{sec4}.

\section{Model setup}
\label{sec2}

In Sec.~\ref{sec21}, we introduce the price impact model with both self-- and cross--impacts, and transform the discrete model into a continuous one. In Sec.~\ref{sec22}, we discuss the cost of trading and derive in detail the function for cross--impact costs. In Sec.~\ref{sec23}, We construct a trading strategy with three free parameters.

\subsection{Trade price}
\label{sec21}

In financial markets, a buy market order will raise or maintain the stock price, while a sell market order will drop or maintain the stock price. The price change on average due to a buyer--initiated or a seller--initiated trade refers to the price impact. This impact can be propagated to the price in a future time. Therefore, the stock price is the result of the accumulation of impacts from all past trades. The impact can be classified as a self--impact and a cross--impact~\cite{Wang2016c}. The self--impact is related to the trades from the impacted stock itself. As the insufficient volumes at the best ask or bid cannot fulfil the large demand of market orders in a short term, leading to a lack of the short--run liquidity, the traded price has to be conceded to a higher ask price or a lower bid price. This instantaneous price change for stock $i$ at time $t$ thus results from the impact of traded volumes $f_i\big(v_i(t)\big)$. However, the price change is not fixed with time~\cite{Bouchaud2004}. When new limit orders come into the order book, the price is reversed towards previous one gradually. Such price change due to the restoration of liquidity in a long term is characterized by a self--impact function $G_{ii}(\tau)$. Differently, the cross--impact~\cite{Wang2016a,Wang2016b,Benzaquen2016} across different stocks is more likely due to the trading information containing the traded volumes and trade signs rather than other information, because the trade of one stock cannot consume the volume of another stock directly showing in the order book. By the transmission of trading information, the stock $j$ has an impact $g_i\big(v_j(t)\big)$ on the stock $i$. With the time increasing, more and more competitive information, such as news, covers the trading information and weakens the impact of stock $j$ gradually. This decaying process is depicted by a cross--impact function $G_{ij}(\tau)$. 

Therefore, taking into account the price impacts from two different stocks $i$ and $j$, the logarithmic midpoint price of stock $i$ at time $t$ can be expressed~\cite{Wang2016c} as
\begin{eqnarray}  \nonumber
\log m_i(t)&=&\sum_{t'<t}\Big[G_{ii}(t-t')f_{i}\big(v_i(t')\big)\varepsilon_i(t')+\eta_{ii}(t')\Big]  \\    \nonumber
		&+&\sum_{t'<t}\Big[G_{ij}(t-t')g_{i}\big(v_j(t')\big)\varepsilon_j(t')+\eta_{ij}(t')\Big]  \\
		&+&\log m_i(-\infty) \ .
\label{eq2.1.1}
\end{eqnarray}
Here, $m_i(t)$ is the midpoint of the best ask price and the best bid price at time $t$. $\varepsilon_i(t)$ and $\varepsilon_j(t)$ are the trade signs of stocks $i$ and $j$, respectively. $\varepsilon_i(t)=+1$ means a buy market order of stock $i$ at time $t$, while $\varepsilon_i(t)=-1$ means a sell market order. $\varepsilon_i(t)=0$ represents a lack of trading or a balance of buy and sell market orders at $t$. The trade signs in Eq.~\eqref{eq2.1.1} clearly indicate the directions of price changes, a buy for price increasing and a sell for price decreasing. Apart from the causes already described by the impact functions, \textit{i.e.}, $f_i\big(v_i(t)\big)$, $g_i\big(v_j(t)\big)$, $G_{ii}(\tau)$,  $G_{ij}(\tau)$, all remaining causes of the price change arising from stock $i$ and stock $j$ are modelled by the random variables $\eta_{ii}(t)$ and $\eta_{ij}(t)$, respectively.

In Eq.~\eqref{eq2.1.1}, the impact functions of traded volumes $f_{i}\big(v_i(t)\big)$ and $g_{i}\big(v_j(t)\big)$ describe the unsigned price changes caused by the unsigned volumes $v_i(t)$ and $v_j(t)$ of market orders. That means buying in and selling out the same volume have the same strength of impact on the stock $i$, but the impact raising or dropping the price is determined by the terms $f_{i}\big(v_i(t)\big)\varepsilon_i(t)$ and $g_{i}\big(v_j(t)\big)\varepsilon_j(t)$. To facilitate the calculation, we merge the trade signs into unsigned volumes and the unsigned impact functions by the following way,
\begin{eqnarray} \nonumber
f_{i}\big(v_i(t)\big)\varepsilon_i(t)&\longrightarrow& \tilde{f}_{i}\big(\nu_i(t)\big) \ , \\
g_{i}\big(v_j(t)\big)\varepsilon_j(t)&\longrightarrow& \tilde{g}_{i}\big(\nu_j(t)\big) \ ,
\label{eq2.1.2}
\end{eqnarray}
where $\tilde{f}_{i}\big(\nu_i(t)\big)$ and $\tilde{g}_{i}\big(\nu_j(t)\big)$ are signed impact functions of signed volumes $\nu_i(t)$ and $\nu_j(t)$. Thus, when selling out the volume $\nu_i(t)$ with $\nu_i(t)>0$, the price either changes $\tilde{f}_{i}\big(-\nu_i(t)\big)$ or changes $-\tilde{f}_{i}\big(\nu_i(t)\big)$, \textit{i.e.}, the negative price change of buying in the same volume. This also meets the case of $\nu_i(t)<0$.  As a result, we have
\begin{equation}
\tilde{f}_{i}\big(-\nu_i(t)\big)=- \tilde{f}_{i}\big(\nu_i(t)\big)\ .
\label{eq2.1.3}
\end{equation}
Analogously, 
\begin{equation}
\tilde{g}_{i}\big(-\nu_j(t)\big)=-\tilde{g}_{i}\big(\nu_j(t)\big) \ .
\label{eq2.1.4}
\end{equation}
With the substitution of Eq.~\eqref{eq2.1.2}, the price impact model~\eqref{eq2.1.1} is revised as 
\begin{eqnarray}  \nonumber
\log m_i(t)&=&\sum_{t'<t}\Big[G_{ii}(t-t')\tilde{f}_{i}\big(\delta\nu_i(t')\big)+\eta_{ii}(t')\Big]  \\    \nonumber
		&+&\sum_{t'<t}\Big[G_{ij}(t-t')\tilde{g}_{i}\big(\delta\nu_j(t')\big)+\eta_{ij}(t')\Big]  \\
		&+&\log m_i(-\infty) \ ,
\label{eq2.1.5}
\end{eqnarray}
where $\delta \nu_i(t')$ and $\delta \nu_j(t')$ are the signed traded volumes in each time interval $\delta t'$. In unit time interval $\delta t''$, the signed traded volumes are the rates of trading,
\begin{eqnarray} \nonumber
\dot{\nu}_i(t'')&=&\frac{\delta \nu_i(t'')}{\delta t''}=\frac{\delta \nu_i(t')}{\delta t'}  \ , \\
\dot{\nu}_j(t'')&=&\frac{\delta \nu_j(t'')}{\delta t''}=\frac{\delta \nu_j(t')}{\delta t'}  \ .
\label{eq2.1.6}
\end{eqnarray}
The positive rates of trading are for buy market orders, while the negative rates for sell market orders. Considering the limit case that $\delta t'\rightarrow \delta t''$, we transform the discrete time process of the price into a continuous process,
\begin{eqnarray}  \nonumber
\log m_i(t)&=&\int_{-\infty}^{t}G_{ii}(t-t')\tilde{f}_{i}\big(\dot{\nu}_i(t')\big)dt'+\int_{-\infty}^{t}\eta_{ii}(t')dt'  \\    \nonumber
		&+&\int_{-\infty}^{t}G_{ij}(t-t')\tilde{g}_{i}\big(\dot{\nu}_j(t')\big)dt'+\int_{-\infty}^{t}\eta_{ij}(t')dt' \\
		&+&\log m_i(-\infty) \ .
\label{eq2.1.7}
\end{eqnarray}
The continuous process of price is on a physical time scale, rather than a trade or an event time scale that considers a trade or an event as a time stamp. For linear impact functions of traded volumes, Eq.~\eqref{eq2.1.7} is hold. For nonlinear ones, however, Eq.~\eqref{eq2.1.7} will return an approximate value for the price.

\subsection{Costs of trading}
\label{sec22}

A trading strategy $\Pi_i=\{\nu_i(t)\}$ is referred to a round--trip trade~\cite{Gatheral2010} of stock $i$ if the total bought--in volume is the same as the total sold--out volume in a trading period $\mathbb{T}_i$, which can be expressed as
\begin{equation}
\int_0^{\mathbb{T}_i} \dot{\nu}_i(t) dt=0 
\label{eq2.2.1}
\end{equation} 
with the rate of trading $\dot{\nu}_i(t)$. During a trading period $\mathbb{T}_i$, the cost of trading is the expected cost of a sequence of small trades,
\begin{equation} 
\Omega_i(\Pi_i)=E\Big[\int_0^{\mathbb{T}_i} \dot{\nu}_i(t)\big(\log m_i(t)-\log m_i(0)\big)dt \Big] \ .
\label{eq2.2.2}
\end{equation}
Here, we use the difference of the logarithmic midpoint prices, \textit{i.e.} the price return, to represent the price change. Due to the trades randomly initiated by buy and sell market orders, the trade price fluctuates between the best ask and the best bid. In contrast, the midpoint price between the best ask and best bid is better to indicate the price trend, as it is raised by a buy market order and lowered by a sell market order. To avoid the dramatic price shifting, the small trades in a round trip are restricted to exchange the volume that is less than the average. The cost of trading in Eq.~\eqref{eq2.2.2} can be separated into the cost induced by the self--impact
\begin{equation}
\Omega_{ii}(\Pi_i)=\int_0^{\mathbb{T}_i} \dot{\nu}_i(t)dt\int_0^t G_{ii}(t-t')\tilde{f}_i\big(\dot{\nu}_i(t')\big)dt' \ ,
\label{eq2.2.3}
\end{equation}
and the cost induced by the cross--impact
\begin{equation}
\Omega_{ij}(\Pi_i)=\int_0^{\mathbb{T}_i} \dot{\nu}_i(t)dt\int_0^t G_{ij}(t-t')\tilde{g}_i\big(\dot{\nu}_j(t')\big)dt' \ . 
\label{eq2.2.4}
\end{equation}
Since the price impacts of all the buy and sell trades in a round trip are averaged to greatly lower the effects of random variables, the costs induced by $\eta_{ii}(t')$ and $\eta_{ij}(t')$ are ignored in above two equations. The total cost of stock $i$ is the sum of the two parts,
\begin{equation}
\Omega_i(\Pi_i)=\Omega_{ii}(\Pi_i)+\Omega_{ij}(\Pi_i) \ .
\label{eq2.2.5}
\end{equation}
For the paired stock $j$ with a trading period $\mathbb{T}_j$, the cost of trading is analogously given by
\begin{equation}
\Omega_j(\Pi_j)=\Omega_{jj}(\Pi_j)+\Omega_{ji}(\Pi_j) \ .
\label{eq2.2.6}
\end{equation}

The two round--trip trades of the stocks $i$ and $j$ start with a time difference $\Delta t$, where to keep the model as simple as possible, we only consider the limit cases, $\Delta t=0$ and $\Delta t\rightarrow \infty$. The former means there is an overlap in the execution periods of the two round--trip trades, leading to the self--impact accompanied with the nontrivial cross--impact. The latter implies the two round--trip trades are executed individually without any overlap in time so that only the self--impact is present in each round--trip trade. If the costs arising from the self--impacts in above two cases are the same, the case with $\Delta t=0$ has an extra cost induced by the cross--impacts,
\begin{equation}
 \Omega_{c}(\Pi_i,\Pi_j)=\Omega_{ij}(\Pi_i)+\Omega_{ji}(\Pi_j) \ .
\label{eq2.2.7}
\end{equation}
Therefore, the cross--impact cost $\Omega_{c}(\Pi_i,\Pi_j)$ determines the optimal execution of the two round-trip trades. If $\Omega_{c}(\Pi_i,\Pi_j)\geqslant0$, executing the two round--trip trades individually with $\Delta t\rightarrow \infty$ is preferred to eliminate the extra cost from the cross--impacts. If $\Omega_{c}(\Pi_i,\Pi_j)<0$, executing the two round--trip trades with $\Delta t=0$ contributes to reduce the trading costs or even to profit from the possible opportunities of arbitrage. The cross--impact cost $\Omega_{c}(\Pi_i,\Pi_j)$ are detailed as follows.

For stock $i$, during the trading period $\mathbb{T}_i$, the volumes are bought in within the first $\theta_i$ period by a rate $\dot{v}_i^{(\textrm{in})}(t)$, and then are sold out totally in the remaining time $(1-\theta_i)\mathbb{T}_i$ by a rate $-\dot{v}_i^{(\textrm{out})}(t)$, where $\theta_i$ is a scaling factor of the bought--in time during the trading period. Analogously for stock $j$ with all the quantities indexed by $j$ instead of $i$. Here, the rates  $\dot{v}_i^{(\textrm{in})}(t)$, $\dot{v}_j^{(\textrm{in})}(t)$, $\dot{v}_i^{(\textrm{out})}(t)$ and $\dot{v}_j^{(\textrm{out})}(t)$ are always positive values. To trace the rates of trading in different time regions, the constant rates are denoted as a function of time $t$ or $t'$ in the following integrals. Furthermore, to reduce the complexity of the integrals, the trading strategies are distributed in the following time regions.

\begin{description}
\item[(I)] $0\leqslant \theta_j\mathbb{T}_j\leqslant \theta_i\mathbb{T}_i\leqslant\mathbb{T}_j$ 
\end{description}

The transformation from buying in to selling out the stock $i$ occurs during the period of stock $j$ being sold out. Thus, the cross--impact costs in Eq.~\eqref{eq2.2.7} can be expanded as
\begin{eqnarray} \nonumber
\Omega_{ij}(\Pi_i)&=&\int_0^{\theta_i\mathbb{T}_i}\dot{v}_i^{(\textrm{in})}(t)dt\int_0^{\theta_j\mathbb{T}_j} G_{ij}(t-t')\tilde{g}_i\big(\dot{v}_j^{(\textrm{in})}(t')\big)dt'  \\ \nonumber
&+&\int_0^{\theta_i\mathbb{T}_i}\dot{v}_i^{(\textrm{in})}(t)dt\int_{\theta_j\mathbb{T}_j}^t G_{ij}(t-t')\tilde{g}_i\big(-\dot{v}_j^{(\textrm{out})}(t')\big)dt'  \\ \nonumber
&+&\int_{\theta_i\mathbb{T}_i}^{\mathbb{T}_i}\big(-\dot{v}_i^{(\textrm{out})}(t)\big)dt\int_0^{\theta_j\mathbb{T}_j} G_{ij}(t-t')\tilde{g}_i\big(\dot{v}_j^{(\textrm{in})}(t')\big)dt'  \\ 
&+&\int_{\theta_i\mathbb{T}_i}^{\mathbb{T}_i}\big(-\dot{v}_i^{(\textrm{out})}(t)\big)dt\int_{\theta_j\mathbb{T}_j}^t G_{ij}(t-t')\tilde{g}_i\big(-\dot{v}_j^{(\textrm{out})}(t')\big)dt'  \ ,
\label{eq2.2.8}
\end{eqnarray}

\begin{eqnarray} \nonumber
\Omega_{ji}(\Pi_j)&=&\int_0^{\theta_j\mathbb{T}_j}\dot{v}_j^{(\textrm{in})}(t)dt\int_0^t G_{ji}(t-t')\tilde{g}_j\big(\dot{v}_i^{(\textrm{in})}(t')\big)dt'  \\ \nonumber
&+&\int_{\theta_j\mathbb{T}_j}^{\mathbb{T}_j}\big(-\dot{v}_j^{(\textrm{out})}(t)\big)dt\int_0^{\theta_i\mathbb{T}_i} G_{ji}(t-t')\tilde{g}_j\big(\dot{v}_i^{(\textrm{in})}(t')\big)dt'  \\ 
&+&\int_{\theta_j\mathbb{T}_j}^{\mathbb{T}_j}\big(-\dot{v}_j^{(\textrm{out})}(t)\big)dt\int_{\theta_i\mathbb{T}_i}^t G_{ji}(t-t')\tilde{g}_j\big(-\dot{v}_i^{(\textrm{out})}(t')\big)dt' \ .
\label{eq2.2.9}
\end{eqnarray}

Here, due to the lag effect of cross--impacts, it is possible that the upper limit of integrals of selling out the volume of one stock goes beyond the trading period of this stock. However, the lag effect after finishing the process of buying in the volume is quickly covered by the effect of selling out the volume, it will not influence on the upper limit of integrals of buying in the volume. The cases in other time regions also have the similar treatment for the intergrals.

\begin{description}
\item[(II)]  $0\leqslant \theta_j\mathbb{T}_j\leqslant\mathbb{T}_j\leqslant \theta_i\mathbb{T}_i $
\end{description}

Before emptying all the bought--in volumes of stock $i$, a round--trip trade of stock $j$ has been fully executed. The cost $\Omega_{ij}(\Pi_i)$ has the same expression as Eq.~\eqref{eq2.2.8}. The cost $\Omega_{ji}(\Pi_j)$ is given by
\begin{eqnarray} \nonumber
\Omega_{ji}(\Pi_j)&=&\int_0^{\theta_j\mathbb{T}_j}\dot{v}_j^{(\textrm{in})}(t)dt\int_0^t G_{ji}(t-t')\tilde{g}_j\big(\dot{v}_i^{(\textrm{in})}(t')\big)dt'  \\ 
&+&\int_{\theta_j\mathbb{T}_j}^{\mathbb{T}_j}\big(-\dot{v}_j^{(\textrm{out})}(t)\big)dt\int_0^t G_{ji}(t-t')\tilde{g}_j\big(\dot{v}_i^{(\textrm{in})}(t')\big)dt'  \ .
\label{eq2.2.10}
\end{eqnarray}

\begin{description}
\item[(III)] $0\leqslant \theta_i\mathbb{T}_i\leqslant \theta_j\mathbb{T}_j\leqslant\mathbb{T}_i$ 
\end{description}

The transformation from buying in to selling out the stock $j$ occurs during the period of emptying all bought--in volumes of stock $i$. Thus, we have the cross--impact costs
\begin{eqnarray} \nonumber
\Omega_{ij}(\Pi_i)&=&\int_0^{\theta_i\mathbb{T}_i}\dot{v}_i^{(\textrm{in})}(t)dt\int_0^t G_{ij}(t-t')\tilde{g}_i\big(\dot{v}_j^{(\textrm{in})}(t')\big)dt'  \\ \nonumber
&+&\int_{\theta_i\mathbb{T}_i}^{\mathbb{T}_i}\big(-\dot{v}_i^{(\textrm{out})}(t)\big)dt\int_0^{\theta_j\mathbb{T}_j} G_{ij}(t-t')\tilde{g}_i\big(\dot{v}_j^{(\textrm{in})}(t')\big)dt'  \\ 
&+&\int_{\theta_i\mathbb{T}_i}^{\mathbb{T}_i}\big(-\dot{v}_i^{(\textrm{out})}(t)\big)dt\int_{\theta_j\mathbb{T}_j}^t G_{ij}(t-t')\tilde{g}_i\big(-\dot{v}_j^{(\textrm{out})}(t')\big)dt' 
\label{eq2.2.11} \ ,
\end{eqnarray}

\begin{eqnarray} \nonumber
\Omega_{ji}(\Pi_j)&=&\int_0^{\theta_j\mathbb{T}_j}\dot{v}_j^{(\textrm{in})}(t)dt\int_0^{\theta_i\mathbb{T}_i}  G_{ji}(t-t')\tilde{g}_j\big(\dot{v}_i^{(\textrm{in})}(t')\big)dt'  \\ \nonumber
&+&\int_0^{\theta_j\mathbb{T}_j}\dot{v}_j^{(\textrm{in})}(t)dt\int_{\theta_i\mathbb{T}_i}^t G_{ji}(t-t')\tilde{g}_j\big(-\dot{v}_i^{(\textrm{out})}(t')\big)dt' \\ \nonumber
&+&\int_{\theta_j\mathbb{T}_j}^{\mathbb{T}_j}\big(-\dot{v}_j^{(\textrm{out})}(t)\big)dt\int_0^{\theta_i\mathbb{T}_i} G_{ji}(t-t')\tilde{g}_j\big(\dot{v}_i^{(\textrm{in})}(t')\big)dt'  \\ 
&+&\int_{\theta_j\mathbb{T}_j}^{\mathbb{T}_j}\big(-\dot{v}_j^{(\textrm{out})}(t)\big)dt\int_{\theta_i\mathbb{T}_i}^t G_{ji}(t-t')\tilde{g}_j\big(-\dot{v}_i^{(\textrm{out})}(t')\big)dt' \ .
\label{eq2.2.12}
\end{eqnarray}

\begin{description}
\item[(IV)] $0\leqslant \theta_i\mathbb{T}_i\leqslant\mathbb{T}_i\leqslant \theta_j\mathbb{T}_j$ 
\end{description}

Before all the bought--in volumes of stock $j$ being sold out, the execution of the round--trip trade of stock $i$ has finished. Thus, the cross--impact cost $\Omega_{ij}(\Pi_i)$ is
\begin{eqnarray} \nonumber
\Omega_{ij}(\Pi_i)&=&\int_0^{\theta_i\mathbb{T}_i}\dot{v}_i^{(\textrm{in})}(t)dt\int_0^t G_{ij}(t-t')\tilde{g}_i\big(\dot{v}_j^{(\textrm{in})}(t')\big)dt'  \\ 
&+&\int_{\theta_i\mathbb{T}_i}^{\mathbb{T}_i}\big(-\dot{v}_i^{(\textrm{out})}(t)\big)dt\int_0^t G_{ij}(t-t')\tilde{g}_i\big(\dot{v}_j^{(\textrm{in})}(t')\big)dt'  \ .
\label{eq2.2.13}
\end{eqnarray}
The expression of the cost $\Omega_{ji}(\Pi_j)$ is the same as Eq.~\eqref{eq2.2.12}.

\subsection{A construction of trading strategies}
\label{sec23}

A round--trip trade ends up when the net volume is zero, leading to 
\begin{equation}
\dot{v}_i^{(\textrm{in})}\theta_i\mathbb{T}_i-\dot{v}_i^{(\textrm{out})}(1-\theta_i)\mathbb{T}_i=0 \ ,
\label{eq2.3.1}
\end{equation}
\begin{equation}
\dot{v}_j^{(\textrm{in})}\theta_j\mathbb{T}_j-\dot{v}_j^{(\textrm{out})}(1-\theta_j)\mathbb{T}_j=0 \ .
\label{eq2.3.2}
\end{equation}
Setting $\dot{v}_i$ and $\dot{v}_j$ as the sums of bought--in rates and sold--out rates for stocks $i$ and $j$, respectively,
\begin{equation}
\dot{v}_i=\dot{v}_i^{(\textrm{in})}+ \dot{v}_i^{(\textrm{out})} \ ,
\label{eq2.3.3}
\end{equation}
\begin{equation}
\dot{v}_j=\dot{v}_j^{(\textrm{in})}+ \dot{v}_j^{(\textrm{out})} \ ,
\label{eq2.3.4}
\end{equation}
the bought--in and sold--out rates can be denoted as
\begin{equation}
\dot{v}_i^{(\textrm{in})}=\kappa_i\dot{v}_i
\qquad \textrm{and} \qquad
\dot{v}_i^{(\textrm{out})}=(1-\kappa_i)\dot{v}_i 
\label{eq2.3.5}
\end{equation}
for stock $i$, and
\begin{equation}
\dot{v}_j^{(\textrm{in})}=\kappa_j\dot{v}_j
\qquad \textrm{and} \qquad
\dot{v}_j^{(\textrm{out})}=(1-\kappa_j)\dot{v}_j 
\label{eq2.3.6}
\end{equation}
for stock $j$, where the scaling factors of the bought--in rates $\kappa_i$ and $\kappa_j$ are bound to
\begin{equation}
0<\kappa_i<1
\qquad \textrm{and} \qquad
0<\kappa_j<1 \ .
\label{eq2.3.7}
\end{equation}
According to Eqs.~\eqref{eq2.3.1}---\eqref{eq2.3.6}, the scaling factors of bought--in time $\theta_i$ and $\theta_j$ can be replaced by 
\begin{equation}
\theta_i=1-\kappa_i 
\qquad \textrm{and} \qquad
\theta_j=1-\kappa_j \ .
\label{eq2.3.8}
\end{equation} 
To connecting the stock $i$ with the stock $j$, we introduce $\zeta_\mathbb{T}$, which links the trading periods of two stocks, 
\begin{equation}
\zeta_\mathbb{T}=\frac{\mathbb{T}_i}{\mathbb{T}_j} \ ,
\label{eq2.3.9}
\end{equation}
and $\zeta_v$, which combines the total bought--in (or sold--out) volumes $v_i$ and $v_j$ of two stocks,
\begin{equation}
\zeta_v=\frac{v_i}{v_j} \ .
\label{eq2.3.10}
\end{equation}
By making use of the definition~\eqref{eq2.1.6}, the sums of bought--in and sold--out rates of the two stocks have the ratio,
\begin{equation} 
\frac{\dot{v}_i}{\dot{v}_j}
=\frac{\zeta_v}{\zeta_{\mathbb{T}}}\frac{(1-\kappa_j)\kappa_j}{(1-\kappa_i)\kappa_i} \ .
\label{eq2.3.11}
\end{equation}

Therefore, to execute two round--trip trades of stocks $i$ and $j$, we need to preset the ratio $\zeta_v$ of the total bought--in volumes, the bought--in rate $\dot{v}_i^{(\textrm{in})}$ and the trading period $\mathbb{T}_i$ according to the practical demand. With three free parameters $\kappa_i$, $\kappa_j$ and $\zeta_{\mathbb{T}}$, we then can work out the remaining quantities by Eqs.~\eqref{eq2.3.3}--\eqref{eq2.3.11}, including the sold--out rate $\dot{v}_i^{(\textrm{out})}$ and  $\dot{v}_j^{(\textrm{out})}$, the bought--in rate $\dot{v}_j^{(\textrm{in})}$, the trading period $\mathbb{T}_j$, and the time for buying in and selling out each stock. As a result, a trading strategy is determined by the set of $\{\kappa_i, \kappa_j,\zeta_{\mathbb{T}}\}$, where the optimal trading strategy is conditioned on the minimal cost of cross--impacts 
\begin{equation}
\Omega_{c}(\Pi_i,\Pi_j)=\textrm{min}\big\{\Omega_c \big(\kappa_i, \kappa_j, \zeta_{\mathbb{T}}\big)\big\} \ .
\label{eq2.3.12}
\end{equation}

\section{Applications to a specific case}
\label{sec3}

The cost functions in Eqs.~\eqref{eq2.2.8}---\eqref{eq2.2.13} contain the impact functions of time lag and of trade volumes. However, these impact functions have not been determined yet. Although Ref.~\cite{Wang2016c} gives the functional form for them, the parameters in the functions depend on the specific stocks. To result in a feasible trading strategy in terms of the cross--impact cost for a specific case, it is necessary to measure the price impacts. Therefore, in Sec.~\ref{sec31}, we introduce the data set used for the empirical measurement. In Sec.~\ref{sec32}, we describe the algorithm for classifying the trade signs, which is crucial for measuring the price impacts. In Sec.~\ref{sec33}, we quantify the impacts of traded volumes, fitted by a power law. In Sec.~\ref{sec34}, with the help of the cross--response functions and the self--correlators of trade signs, we measure the cross--impacts of time lag between two stocks. In Sec.~\ref{sec35}, using the fitted and preset parameters, we carry out and discuss the trading strategies with respect to the cross--impact costs. 

\subsection{Data sets}
\label{sec31}

We apply our trading strategy to a specific pair of stocks, Apple Inc. (AAPL) and Microsoft Corp. (MSFT), where AAPL is indexed by $i$ and MSFT is indexed by $j$ in the following. We use the Trades and Quotes (TAQ) data set, where the data of two stocks comes from the NASDAQ stock market in 2008. For a given stock in each year, the TAQ data set contains a trade file recording all the information of each trade and a quote file recording all the information of each quote. The information of trades and quotes has the resolution of one second. However, more than one trade or quote may be found  in TAQ data set on the time scale smaller than one second. In addition, we only consider the trading days that AAPL and MSFT all have trades so as to have the cross--impacts during the intraday trading time. To avoid the dramatic fluctuation of prices due to any artifact at the opening and closing of the market, we exclude the data in the first and the last ten minutes of trading.

\subsection{Trade signs}
\label{sec32}

The trade sign plays a crucial role in measuring the price impacts from empirical data. Since the TAQ data set lacks of the information about the trade type (buy or sell) or the trade sign, a method to identify the trade signs is required. One representative algorithm put forward by Lee and Ready~\cite{Lee1991} is to compare the trade price with the preceding midpoint price. However, it is difficult to employ this algorithm to identify the signs for the trades during the one--second interval, because we cannot match those trades with their preceding midpoint prices without a higher resolution of TAQ data set. In view of this, we resort to the algorithm described in our previous study~\cite{Wang2016a}. The sign $\varepsilon(t;n)$ of $n$-th trade in the time interval $t$ results from the sign of price change if the prices $S(t;n)$ and $S(t;n-1)$ for two consecutive trades are different, or otherwise from the preceding trade sign,
\begin{eqnarray}       
\varepsilon(t;n)=\left\{                  
\begin{array}{lll}    
\mathrm{sgn}\bigl(S(t;n)-S(t;n-1)\bigr)  & \ ,  & \quad \mbox{if} \quad S(t;n)\neq S(t;n-1) \ , \\    
\varepsilon(t;n-1) & \ ,  & \quad \mbox{otherwise} \ .
\end{array}           
\right.    
\label{eq3.2.1}           
\end{eqnarray}
It is worth to mention that the trade price $S(t;n)$, found directly from the trade file of TAQ data set, differs from the midpoint price $m(t)$, which is obtained from the last quote prior to the time interval $t$ in the quote file. Moreover, the trade sign $\varepsilon(t)$ for the time interval of one second is defined as 
\begin{eqnarray}       
\varepsilon(t)=\left\{                  
\begin{array}{lll}    
\mathrm{sgn}
\left(\sum\limits_{n=1}^{N(t)}\varepsilon(t;n)\right) & \ , \qquad & \mbox{if} \quad N(t)>0 \ , \\    
                                                0 & \ , \qquad & \mbox{if} \quad N(t)= 0 \ .
\end{array}           
\right.    
\label{eq3.2.2}              
\end{eqnarray}
That is a sign function of the sum of the trade signs $\varepsilon(t;n)$ in time interval $t$ if there were trades in this interval. Otherwise, the absence of trading in $t$ leads $\varepsilon(t)$ to be zero. Same as the $\varepsilon(t)$ in Eq.~\eqref{eq2.1.1}, the sign $\varepsilon(t)$ here indicates the trade type of market orders. $\varepsilon(t)=+1$ ($\varepsilon(t)=-1$) means a majority of buy (sell) market orders in time interval $t$, and $\varepsilon(t)=0$ means a lack of trading or a balance of buy and sell market orders in this interval. The tests of this algorithm using the TotalView--ITCH data set, carried out in Ref.~\cite{Wang2016a}, reveals the average accuracy of $85\%$ for Eq.~\eqref{eq3.2.1} and of $82\%$ for Eq.~\eqref{eq3.2.2} to identify the trade signs.

\subsection{Measurement for impacts of traded volumes}
\label{sec33}

The traded volume in this study refers to the aggregation of all the traded volumes in the time interval $t$. To put different stocks in the same footing, the traded volumes of each stock are normalized by dividing the average of traded volumes over a whole year,
\begin{equation}
v(t)=\frac{T\sum_{n=1}^{N(t)}v(t;n)}{\sum_{t=1}^{T}\sum_{n=1}^{N(t)}v(t;n)} \ ,
\label{eq3.3.1}
\end{equation}
where $v(t;n)$ is the volume of the $n$-th trade in the time interval $t$, $N(t)$ is the number of trades in $t$, and $T$ is the total time intervals for trading during a whole year. Thus, $v(t)<1$ indicates that the traded volumes are smaller than their average. Conditioned on the unsigned volumes $v_j(t)$, the price change of stock $i$, on average, due to the trades of stock $j$, \textit{i.e.}, the price cross--response, is given~\cite{Potters2003,Wang2016c} by
\begin{equation}
R_{ij}(v_j, \tau)=\Big\langle r_{i}(t,\tau)\varepsilon_j(t)\Big|v_j(t)\Big\rangle _t \ ,
\label{eq3.3.2}
\end{equation}
where $\langle \cdots \rangle_t$ means the average over all the time $t$, and the price change $r_{i}(t,\tau)$ at time $t$ with a time lag $\tau$ is defined as the difference of logarithmic midpoint prices,
\begin{equation}
r_i(t,\tau) =\log m_i(t+\tau) -\log m_i(t) =\log\frac{m_i(t+\tau)}{m_i(t)} \ .
\label{eq3.3.3}
\end{equation}
Since the influence of traded volumes is independent of the time lag, Eq.~\eqref{eq3.3.2} can be approximately decomposed into
\begin{equation}
R_{ij}(v_j, \tau) \approx R_{ij}(\tau) g_i(v_j) \ ,
\label{eq3.3.4}
\end{equation}
where
\begin{equation}
R_{ij}(\tau)=\Big\langle r_{i}(t,\tau)\varepsilon_j(t)\Big\rangle _t \ 
\label{eq3.3.5}
\end{equation}
is the price cross--response depending on the time lag, and $g_i(v_j)$ is the impact function of traded volumes. For the average price change of stock $j$ induced by stock $i$, analogously we have, 
\begin{equation}
R_{ji}(v_i, \tau) \approx R_{ji}(\tau) g_j(v_i) \ .
\label{eq3.3.6}
\end{equation}

\begin{figure}[tb]
  \begin{center}
    \includegraphics[width=1\textwidth]{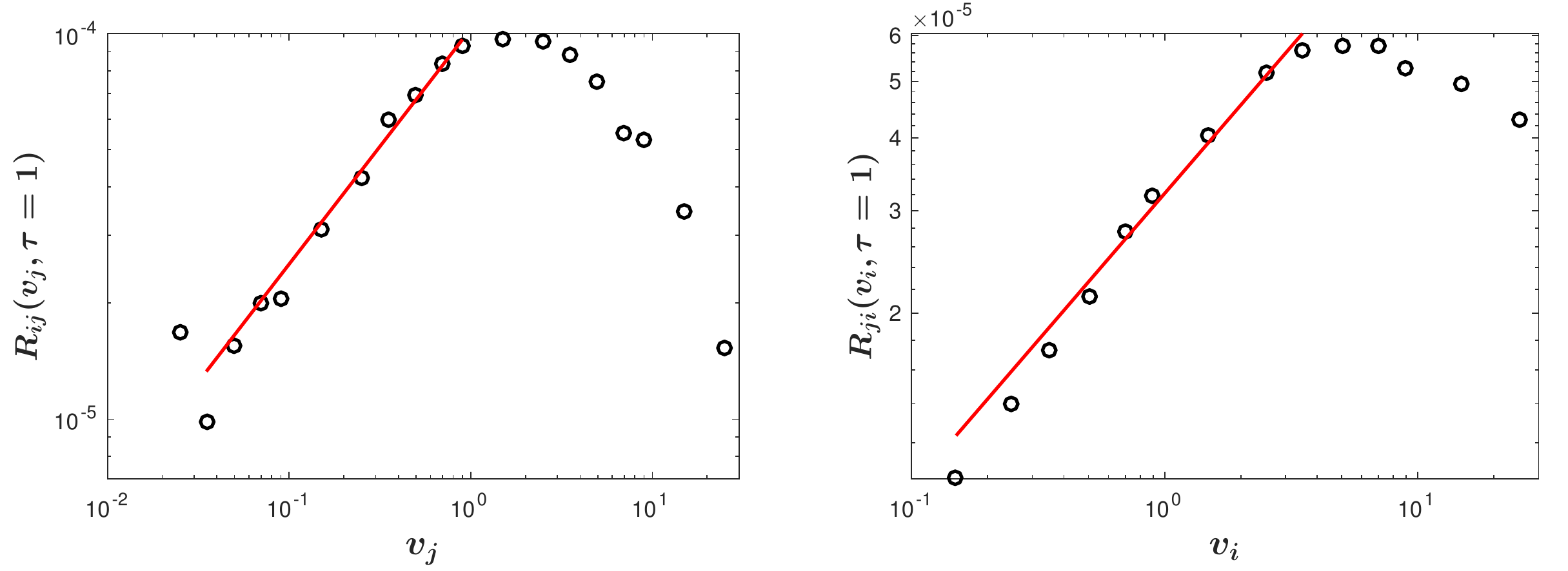} 
  \end{center}
   \setlength{\abovecaptionskip}{-0.3cm} 
\caption{Empirical (circle) and fitted (line) results of impact functions of traded volumes. Stock $i$ is AAPL, and stock $j$ is MSFT.}
 \label{Fig.1}
\end{figure}

Using the empirical data of AAPL and MSFT, we carry out the dependence of price changes on the traded volumes with $\tau=1$, as shown in Fig.~\ref{Fig.1}. Coinciding with Ref.~\cite{Wang2016c}, the dependencies for small traded volumes are fitted well by a power law, 
\begin{equation}
g_i(v_j)=v_j^{\delta_{ij}}
\qquad \textrm{and}\qquad
g_j(v_i)=v_i^{\delta_{ji}} \ ,
\label{eq3.3.7}
\end{equation}  
where the parameters $\delta_{ij}$ and $\delta_{ji}$ for AAPL and MSFT, respectively, are listed in Table~\ref{tab1}. To make the trading strategy feasible, we limit the volume of each trade in strategies to be smaller than the average. 

We notice that the traded volumes and the impact functions in Eq.~\eqref{eq3.3.7} are all unsigned. With the positive rates of trading to buy a stock, the signed impact functions of traded volumes in Eqs.~\eqref{eq2.2.8}---\eqref{eq2.2.13} are the same as the unsigned ones, as shown in Eq.~\eqref{eq3.3.7}. With the negative rates of trading to sell a stock, according to Eq.~\eqref{eq2.1.4}, the signed impact functions turn to unsigned ones by the following way,
\begin{eqnarray} 
&\tilde{g}_i\big(-\dot{v}_j^{(\textrm{out})}(t')\big)=-\tilde{g}_i\big(\dot{v}_j^{(\textrm{out})}(t')\big)=-g_i\big(\dot{v}_j^{(\textrm{out})}(t')\big) \ ,\\ 
&\tilde{g}_j\big(-\dot{v}_i^{(\textrm{out})}(t')\big)=-\tilde{g}_j\big(\dot{v}_i^{(\textrm{out})}(t')\big)=-g_j\big(\dot{v}_i^{(\textrm{out})}(t')\big) \ .
\label{eq3.3.8}
\end{eqnarray}

\begin{table}[bp]
\renewcommand\arraystretch{1.5}
\caption{Parameters for impact functions}
\begin{center}
\begin{tabular}{@{\hskip 0pt}c@{\hskip 15pt}c@{\hskip 10pt}c@{\hskip 10pt}c@{\hskip 10pt}c@{\hskip 10pt}c@{\hskip 10pt}c@{\hskip 15pt}c@{\hskip 15pt}c@{\hskip 15pt}c@{\hskip 0pt}} 
\hline 
$\delta_{ij}$	& $\delta_{ji}$ &$\langle g_i(v_j(t))\rangle_t$ &$\langle g_j(v_i(t))\rangle_t$ & $\Gamma_{0,ij}$& $\Gamma_{0,ji}$  & $\tau_{0,ij}$ &  $\tau_{0,ji}$ &  $\beta_{ij}$ &  $\beta_{ji}$\\ 
\hline
0.61& 0.50	& 0.40	&0.60	& 1.13 $\times10^{-4}$ &0.79$\times10^{-4}$ & 7.34 & 4.75 &0.14 & 0.03	\\
\hline
\end{tabular}
\end{center}
\label{tab1}
\end{table}

\subsection{Measurement for cross--impacts of time lag}
\label{sec34}

\begin{figure}[tb]
  \begin{center}
    \includegraphics[width=1\textwidth]{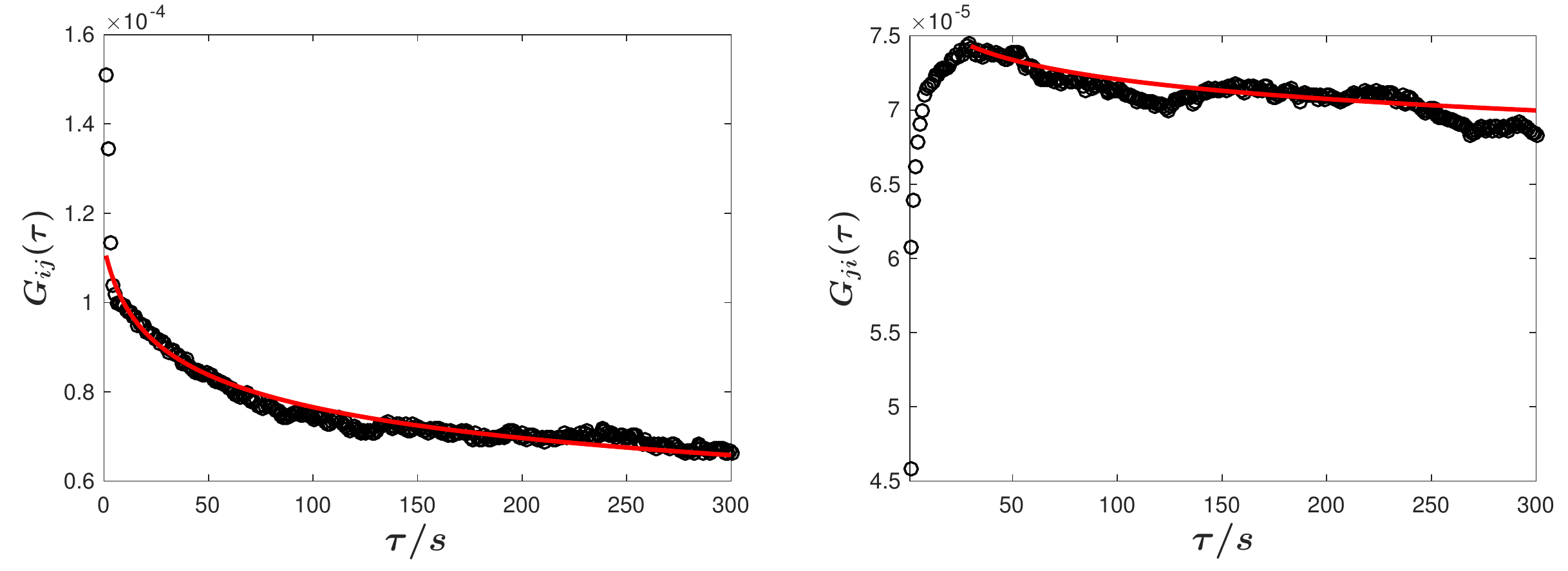} 
  \end{center}
   \setlength{\abovecaptionskip}{-0.3cm} 
\caption{Empirical (circle) and fitted (line) results of impact functions of time lag. Stock $i$ is AAPL, and stock $j$ is MSFT.}
 \label{Fig.2}
\end{figure}

The price cross--response comprises two components~\cite{Wang2016c}. One arises from the self--impacts and is related to the cross--correlators of trade signs. The other one results from the cross--impacts and is related to the self--correlators of trade signs. Here, we focus on the response component containing the cross--impacts, which is given by 
\begin{eqnarray} \nonumber
R_{ij}^{(S)}(\tau)&=&\sum_{0\leq t<\tau} G_{ij}(\tau-t)\left\langle g_i\big(v_j(t)\big)\right\rangle_t \Theta_{jj}(t) \\
		&&+\sum_{t<0}\Big[ G_{ij}(\tau-t)-G_{ij}(-t)\Big]\left\langle g_i\big(v_j(t)\big)\right\rangle_t \Theta_{jj}(-t) \ .
\label{eq3.4.1}
\end{eqnarray}
The superscript $(S)$ in the response function indicates the response component related to the the self--correlator $\Theta_{jj}(\tau)$, defined as
\begin{equation}
\Theta_{jj}(\tau)=\Big\langle\varepsilon_j(t+\tau)\varepsilon_j(t)\Big\rangle_t \ .
\label{eq3.4.2}
\end{equation}
 In Eq.~\eqref{eq3.4.1}, by replacing $\tau-t$ or $-t$ with $\tau'$  in each cross--impact function of time lag,
\begin{eqnarray} \nonumber
R_{ij}^{(S)}(\tau)&=&\sum_{0<\tau'\leq\tau}G_{ij}(\tau')\left\langle g_i\big(v_j(t)\big)\right\rangle_t \Theta_{jj}(\tau-\tau') \\ \nonumber
			 &&+\sum_{\tau'>\tau} G_{ij}(\tau')\left\langle g_i\big(v_j(t)\big)\right\rangle_t \Theta_{jj}(-\tau+\tau') \\
			 &&-\sum_{\tau'>0} G_{ij}(\tau')\left\langle g_i\big(v_j(t)\big)\right\rangle_t \Theta_{jj}(\tau') \ ,
\label{eq3.4.3}
\end{eqnarray}
and using the symmetric property of sign self--correlators $\Theta_{jj}(\tau)=\Theta_{jj}(-\tau) $, we have
\begin{equation}
\frac{R_{ij}^{(S)}(\tau)}{\left\langle g_i\big(v_j(t)\big)\right\rangle_t}=\sum_{\tau'=1}^{\infty}A_{jj}(\tau,\tau')G_{ij}(\tau') \ ,
\label{eq3.4.4}
\end{equation}
where
\begin{equation}
A_{jj}(\tau, \tau')=\Theta_{jj}(\tau-\tau')- \Theta_{jj}(\tau') \ .
\label{eq3.4.5}
\end{equation}
The component $R_{ij}^{(S)}(\tau)$ is the cross--response $R_{ij}(\tau)$ weighted by a quantity $w_i$ with $0<w_i<1$. Therefore, the cross--impact of time lag entering the impact matrix $G_{ij}$ can be quantified from empirical data by 
\begin{equation}
G_{ij}=\frac{w_i}{\left\langle g_i\big(v_j(t)\big)\right\rangle_t}A_{jj}^{-1}R_{ij} \ ,
\label{eq3.4.6}
\end{equation}
where $A_{jj}$ is the matrix of sign correlators with the elements worked out by Eq.~\eqref{eq3.4.5} and $R_{ij}$ is the response matrix with the elements $R_{ij}(\tau)$, defined by Eq.~\eqref{eq3.3.5}.
Analogously for stock $j$, we have
\begin{equation}
G_{ji}=\frac{w_j}{\left\langle g_j\big(v_i(t)\big)\right\rangle_t}A_{ii}^{-1}R_{ji}\ .
\label{eq3.4.7}
\end{equation}
Although we can estimate the weight $w_i$ and $w_j$ by a complicated method, as introduced in Ref.~\cite{Wang2016c}, to facilitate the calculation, we assume $w_i=w_j$ and further normalize the cross--impact of time lag by $w_i$ or $w_j$. By this way, it also normalizes the cost of trading according to Eqs.~\eqref{eq2.2.8}---\eqref{eq2.2.13}, but it does not change the sign of the cost, used to distinguish the profit from the cost. 

Using Eqs.~\eqref{eq3.4.6} and \eqref{eq3.4.7}, we work out the empirical cross--impacts of time lag between AAPL and MSFT, shown in Fig.~\ref{Fig.2} with circles. To obtain the cross--impacts in the first 300 seconds, we replace the $\infty$ in Eq.~\eqref{eq3.4.4} by a large cut--off of 3000 seconds. Due to the fluctuations of sign self--correlators and of cross-responses, the cross--impacts in small time lags are unstable. We thus extract the empirical results with stably decaying for parameter fits. To fit to empirical data, here, we employ simplified power--law functions instead of the complicated functional form in Ref.~\cite{Wang2016c}, 
\begin{equation}
G_{ij}(\tau)=\frac{\Gamma_{0,ij}}{\Big(1+\frac{\tau}{\tau_{0,ij}}\Big)^{\beta_{ij}}} 
\qquad \textrm{and} \qquad
G_{ji}(\tau)=\frac{\Gamma_{0,ji}}{\Big(1+\frac{\tau}{\tau_{0,ji}}\Big)^{\beta_{ji}}}  \ ,
\label{eq3.4.8}
\end{equation}
where $\tau_{0,ij}$ and $\tau_{0,ji}$ are the time scales having the positive values, $\beta_{ij}$ and $\beta_{ji}$ are the rates of decaying, and $\Gamma_{0,ij}$ and $\Gamma_{0,ji}$ are the temporary impact components per share. The fitted values of these parameters are listed in Table~\ref{tab1}.

\subsection{Computations and discussions of trading strategies}
\label{sec35}

To obtain the trading strategy, we consider to totally buy in the same volume for AAPL and MSFT, such that $\zeta_v=1$. Further, we set the trading period of AAPL as $\mathbb{T}_i=1$ unit of time. For one unit of time, we plan to buy in 0.1 times average traded volume of AAPL, resulting in $\dot{v}_i^{(\textrm{in})}=0.1$. With these preset values and fitted parameters listed in Table~\ref{tab1}, we carry out the trading strategies $\{\kappa_i, \kappa_j,\zeta_{\mathbb{T}}\}$ in four time regions using the cost function~\eqref{eq2.2.7}. The four time regions lead to the three free parameters in the strategies bound to the conditions,
\begin{eqnarray}
\left\{                  
\renewcommand\arraystretch{1.7}  
\begin{array}{lcl}
\frac{1-\kappa_j}{1-\kappa_i}\leqslant\zeta_{\mathbb{T}}\leqslant\frac{1}{1-\kappa_i} && (\textrm{region I} ) \ ,\\
\zeta_{\mathbb{T}}\geqslant\frac{1}{1-\kappa_i}&& (\textrm{region II} ) \ ,\\
1-\kappa_j\leqslant\zeta_{\mathbb{T}}\leqslant\frac{1-\kappa_j}{1-\kappa_i}&& (\textrm{region III})\ , \\
0<\zeta_{\mathbb{T}}\leqslant1-\kappa_j&& (\textrm{region IV}) \ .
\end{array}           
\right.    
\label{eq3.5.1}           
\end{eqnarray}
Due to the boundary conditions, with a given ratio of trading periods $\zeta_{\mathbb{T}}$, one may not obtain the numerical solution for the cost function~\eqref{eq2.2.7}. 

\begin{figure}[htbp]
  \begin{center}
    \includegraphics[width=0.8\textwidth]{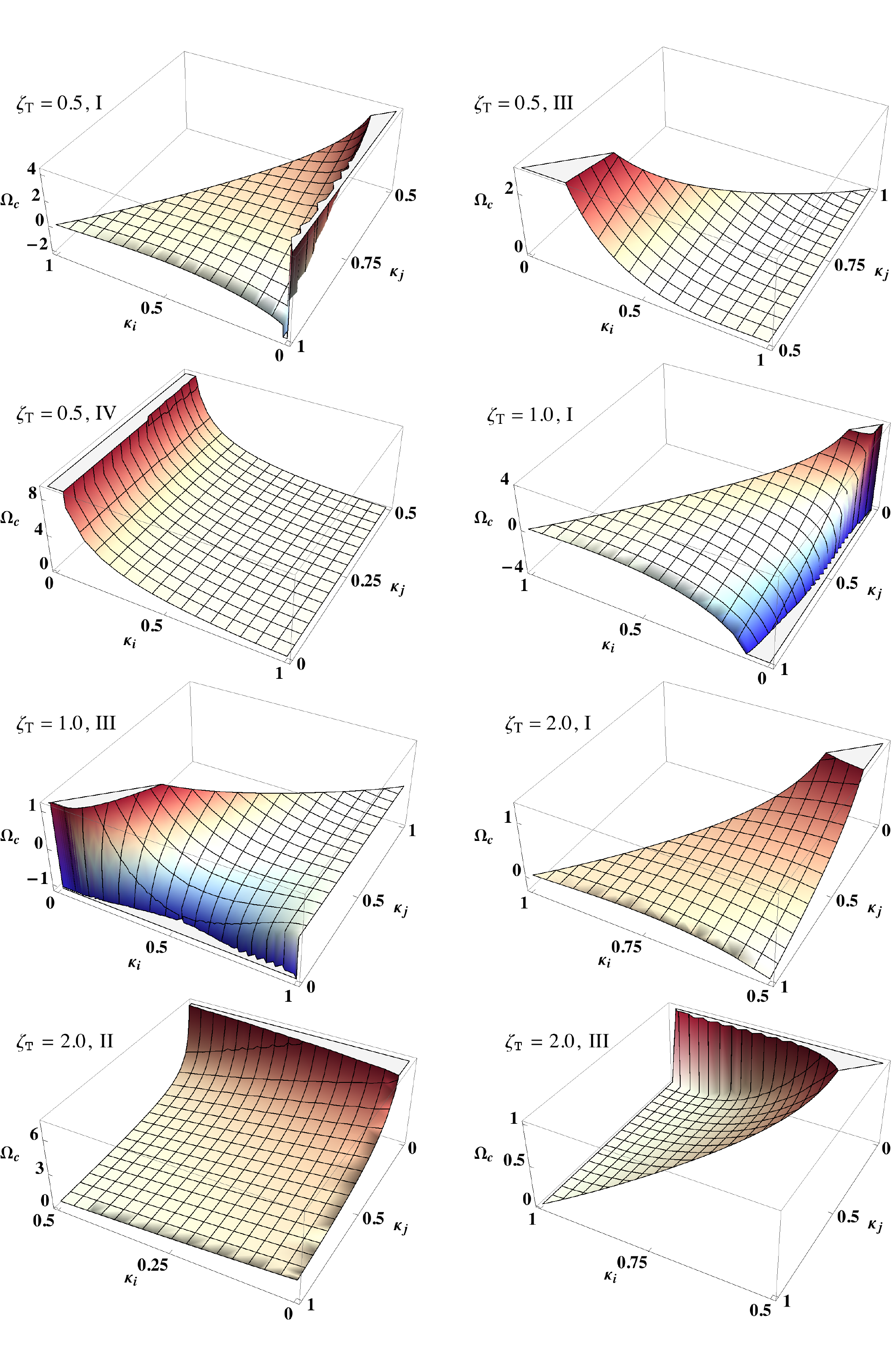} 
  \end{center}
   \setlength{\abovecaptionskip}{-0.7cm} 
\caption{Trading strategies with respect to the costs $\Omega_c$. The increasing of positive and negative costs is displayed by the colour from white to dark red and to dark blue, respectively. Zero cost is indicated by white. To view clearly, the directions and ranges of axes are adjusted for specific cases. The costs of trading $\Omega_c$ are all rescaled by multiplying $10^6$. }
 \label{Fig.3}
\end{figure}

As examples, here, we consider the ratio of trading periods $\zeta_{\mathbb{T}}=0.5$, 1, and 2, meaning the trading period of AAPL is the half of, the same as, and the twice of the trading period of MSFT, respectively. Leaving out the cases without numerical solutions, the cross--impact costs depending on the scaling factors of bought--in rates $\kappa_i$ and $\kappa_j$ are displayed in Fig.~\ref{Fig.3}. In terms of the costs, the trading strategies can be classified as two types, one with the non--negative cost and the other one with the negative cost. For the non--negative cost, it is better to execute the two round--trip trades individually without any overlap in time, \textit{i.e.} $\Delta t\rightarrow\infty$ so as to circumvent the extra cost. If the two round--trip trades inevitably start at the same time, by using the strategy $\{\kappa_i, \kappa_j, \zeta_{\mathbb{T}}\}$ with the minimal cross--impact costs, one can lower the total cost of trading. Taking $\zeta_{\mathbb{T}}=2$ as an example, for the two round--trip trades starting at the same time, we find that the minimal positive cost in region III is at the position of the maximal $\kappa_i$ and $\kappa_j$. It suggests quickly buying in AAPL and MSFT and then slowly selling out them can lower the  total cost of trading to some extent. On the other hand, the presence of the negative cost is possible, especially at a small time scale when the market has not reached to an efficient state~\cite{Wang2016c}. Such case can be seen when $\zeta_{\mathbb{T}}=1$ in regions I and III. The negative cost of trading implies the possible opportunities of arbitrage or a reduction of the total cost for trading. In particular, by minimizing the cross--impact cost to obtain an optimal execution strategy $\{\kappa_i, \kappa_j, \zeta_{\mathbb{T}}\}$, one can maximize the possibility of arbitrage.

\section{Conclusions}
\label{sec4}

We extend the framework of trading strategies for single stocks~\cite{Gatheral2010} to a pair of stocks. For one stock, to lower the execution cost from price self--impacts, traders favour to submit a sequence of small trades. A round trip for buying in and selling out a sequence of small trades is termed a round--trip trade. By considering the executions of two round--trip trades from different stocks, we construct a trading strategy $\{\kappa_i, \kappa_j,\zeta_{\mathbb{T}}\}$, which can be described by the trading rates $\kappa_i$ and $\kappa_j$ of the paired stocks and the ratio of their trading periods $\zeta_{\mathbb{T}}$. By minimizing the cross--impact cost, one can obtain the optimal execution strategy for the two round--trip trades. 

We apply our trading strategy to a pair of stocks, AAPL and MSFT. To determine the impact functions in the strategy, we measure the cross--impacts of time lag and of traded volumes using the empirical data. By numerical computation with the fitted parameters and the preset values, we picture the trading strategy in terms of the cross--impact cost. The positive cost suggests that the individual executions of two round--trip trades without any overlap in time can circumvent the extra cost. The negative cost implies that the two round--trip trades starting at the same time lead to the possible opportunities of arbitrage or a reduction of the total cost for trading. The different ways for order execution reveal the influence of cross--impacts on the optimal trading strategy. Certainly, an improved strategy with respect to the cross--impact cost is called for, but it is beyond this study.

\section*{Acknowledgements}
We thank T. Guhr, D. Waltner, S. M. Krause and M. Akila for helpful discussions. S. Wang acknowledges the financial support from the China Scholarship Council (Grant No. 201306890014). 


\end{document}